\documentstyle[12pt]{article}

\begin{document}
\topmargin 0pt
\oddsidemargin 5mm

\setcounter{page}{1}

\vspace{2cm}
\begin{center}

{APPROXIMATE ANALYTICAL DESCRIPTION OF THE UNDERDENSE \\
SHORT PLASMA LENS}

{\large A.Ts. Amatuni}\\
{\em Yerevan Physics Institute}\\
{Alikhanian Brother's St. 2, Yerevan 375036, Republic of Armenia}

\end{center}

\vspace {5mm}
\centerline{{\bf{Abstract}}}

{The perturbative approach for describing the underdense 
plasma--ultrarelativistic
electron bunch system is developed,using the ratio $\frac{n_0}{n_b}$ as a small
parameter ($n_b$-bunch,$n_0$-plasma electron densities). Focusing of the 
electron bunch emerged in the first approximation of the perturbative procedure
as a result of the plasma electrons redistribution.Focusing gradient and 
strength for ultrarelativistic,flat,uniform and short bunch are obtained and
compared with the previous results.}

keywords:{plasma focusing,underdense plasma lens,perturbative approach}

\section{\small{INTRODUCTION}}

Plasma focusing devices,being compact,simple and effective elements are
promissing for obtaining electron (positron) beams of very small spot size,
requested for future high energy linear colliders.Theoretical predictions
and investigations of the plasma lenses in overdense ($n_b<n_0 \equiv n_p$) and
underdense ($n_b>n_0,n_b$-bunch electron density,$n_0$-cold neitral plasma 
electron
density),performed last decade \cite{A}-\cite{H} and were supplemented by 
experimental tests,carried out at ANL \cite{I}-\cite{J},Tokio University-KEK
\cite{K} and UCLA \cite{L} for overdense plasma lens regime.

The theoretical treatment of the overdense plasma lens,performed in linear
approximation,using ratio $\frac{n_b}{n_0}$ as a small parameter,is more or less
complete,at least in the frame of the cold plasma and rigid electron bunch
approximations.

The existing theoretical approaches to underdense plasma lens are 
phenomenological by nature.

It seems,that more detailed description of the underdense plasma lens is 
needed,in particular,
taking into account the program for the future experimental investigations
\cite{M},which include underdense regime too.

Using the ratio $\frac{n_0}{n_b} \ll 1$ as a small parameter,and developing the
subsequent perturbative approach for the description of the underdense plasma
lens,it is possible to achieve,at least the same level of the understanding
of the focusing phenomenon in underdense regime as that of overdense
plasma lens case.

The present work is devoted to this goal.In the next section the perturbative 
approach is developed for underdense plasma regime for the system of the 
Maxwell equations and hydrodinamical equations of the motion of the plasma 
electrons (ions are assumed immobile).The third section devoted to zero 
approximation calculations of the generated fields and plasma electron 
velocities,
assuming that bunch is ultrarelativistic and short enough.In the section 4 
the first
order approximation results for plasma electrons redistribution and transverse
force (sum of the electric and magnetic forces) are given.Section 5 contents
the disscusion of the obtained results.In particular,it is mentioned that 
physical picture of focusing phenomenon in underdense case is based on the 
redistribution of the plasma electrons around the driving bunch.The effect
of the ion column in ultrarelativistic case is negligible,compared to the
focusing due to  redistributed plasma electrons.

The mathematical technique developed and above mentioned physical discription of
the focusing phenomenon may be used for more detailed,coherent and associated
with the experimental program \cite{M} computer calculations.

\section{\small{PERTURBATIVE APPROACH FOR UNDERDENSE 
        PLASMA LENS}}

For definiteness,consider the flat electron bunch with the vertical dimension
$2b$,which is assumed much smaller than horizontal dimension $2a$;
longitudinal dimensions,which are arbitrary,are $2d$.The bunches of such a 
geometry are suitable for future high energy linear electron-positron colliders
\cite{M}.

Bunch electrons uniform density is $n_b$ and bunch is considered as 
ultrarelativistic and rigid one.Plasma electrons density is $n_0 \ll n_b$,plasma
is neutral,cold and with the immobile ions.

The geometry of the bunch,which moves in lab system through the plasma with 
constant velocity  $v_0$,allows to consider the electric and magnetic fields 
components as follows:

$E_x=0,E_y,E_z \neq 0;B_x \neq 0,B_y=B_z=0$

This more or less evident approximation for the flat beams with large aspect 
ratio,along with the
condition $\frac{\partial}{\partial{x}} \ll \frac{\partial}{\partial{y}}$ was
introduced in \cite{N1}.In \cite{O1} it was pointed out, that the assumption
$B_y \ll B_x \equiv B$ "is not universally true everywhere inside the 
even very flat beam".It is neccessary to take this into account at precise 
analytical or computer calculations.

All physical quantities of the problem are considered as a functions of the
arguments $y$ and $\tilde{z}=z-v_0t$ only (steady state regime).Introduce
the dimensionless arguments ${y'}_1\tilde{z'}=k_by,k_b\tilde{z};t'={\omega}_bt,
{\omega}_b^2=
\frac{4\pi{e}^2n_b}{m},k_b=\frac{{\omega}_b}{c}$,and dimensionless variables
${E'}_{y,z}={(\frac{{\omega}_bmc}{e})}^{-1}E_{y,z};B'={(\frac{{\omega}_bmc}
{e})}^{-1}B_x$;

Followig \cite{N,O,H}define the generalized plasma electron velocities as

$V_y=\frac{v_{ey}}{v_0-v_{ez}}=\frac{{\beta}_{ey}}{\beta-{\beta}_{ez}},
\beta=\frac{v_0}{c},{\beta}_{ez,y}=\frac{v_{e,zy}}c, \
V_z=\frac{v_{ez}}{v_0-v_{ez}}=\frac{{\beta}_{ez}}{\beta-{\beta}_{ez}}$

and generalized plasma electron density as

$N=\frac{n_e}{n_b}(1-\frac{v_{ez}}{v_0})=\frac{n_e}{n_b}(1-\frac{{\beta}_{ez}}
{\beta}),
\frac{n_e}{n_b}=\frac{{\beta}N}{\beta-{\beta}_{ez}}=N(1+V_z)$,

(BTFCh transformations \cite{N}),it is possible to rewrite the continuity 
equation in the following form:

\begin{equation}
\label{AA}
-\frac{\partial{N}}{\partial{z}}+\frac{\partial{(NV_y)}}{\partial{y}}=0
\end{equation}

(in (\ref{AA}) and what follows the tilda and prime superscripts are removed).
The Maxwell equations in this case are:

\begin{eqnarray}
\label{AB}
\begin{array}{l}
1.\qquad \displaystyle \frac{\partial{B}}{\partial{y}}=\beta+{\beta}NV_z+{\beta}
\frac{\partial{E}}{\partial{z}},\\ 
2.\qquad \displaystyle \frac{\partial}{\partial{z}}(B+{\beta}E_y)=-{\beta}NV_y,
\\ 
3.\qquad \displaystyle \frac{\partial}{\partial{z}}({\beta}B+E_y)=
\frac{\partial{E_z}}{\partial{y}},\\ 
4.\qquad \displaystyle \frac{\partial{E_z}}{\partial{z}}+\frac{\partial{E_y}}
{\partial{y}}=-[1-n_0/n_b+N(1+V_z)],
\end{array}
\end{eqnarray}

Eqs. (\ref{AB}) are valid for $-d \leq z \leq d,-b \leq y \leq b$;when $|z|>d,
|y|>b$ in right hand side of eqs. (\ref{AB}.1) and  (\ref{AB}.4) the quantities
$\beta$ and $-1$ subsequently are absent.Eq. (\ref{AB}.4) follows from 
({\ref{AA}), (\ref{AB}.1),(\ref{AB}.2).

The eqs. of the motion are:

\begin{eqnarray}
\label{AC}
\begin{array}{l}
1.\qquad \displaystyle {\dot{V}}_z=\frac{V^{1/2}}{{\beta}^2}
\left[-E_z(1+2V_z+\frac{V_z^2}{{\gamma}^2})+{\beta}^2V_yV_z
\left(E_y+\frac{1}{\beta}B\right)+{\beta}V_yB\right], \\
2.\qquad \displaystyle {\dot{V}}_y=\frac{V^{1/2}}{{\beta}^2}\left[-E_y(1+V_z-
{\beta}^2V_y^2)-E_zV_y(1+\frac{V_z}{{\gamma}^2})-\beta(V_z-V_y^2)B\right]
\end{array}
\end{eqnarray}

where 

\begin{equation}
\label{AD}
V \equiv 1+2V_z+\frac{V_z^2}{{\gamma}^2}-{\beta}^2V_z^2
\end{equation}

\begin{equation}
\label{AE}
{\dot{V}}_y \equiv \frac{\partial{V_y}}{\partial{z}}+V_y\frac{\partial{V_y}}
{\partial{y}},{\dot{V}}_z \equiv -\frac{\partial{V_z}}{\partial{z}}+V_y
\frac{\partial{V_z}}{\partial{y}};
\end{equation}

Adopting the condition $\frac{n_o}{n_b} \ll 1$,decompose the quantities in
question in the following series:

\begin{eqnarray}
\label{AF}
N={\epsilon}N_1+{\epsilon}^2N_2+... \\ \nonumber
V_z=V_{z0}+{\epsilon}V_{z1}+{\epsilon}^2V_{z2}+... \\ \nonumber
V_y=V_{y0}+{\epsilon}V_{y1}+{\epsilon}^2V_{y2}+... \\ \nonumber
E_{y,z}=E_{y,z0}+{\epsilon}E_{y,z1}+{\epsilon}^2E_{y,z2}+... \\ \nonumber
B=B_0+{\epsilon}B_1+{\epsilon}^2B_2+...
\end{eqnarray}

where $\epsilon=\frac{n_0}{n_b}$ (in what follows ${\epsilon}^n$ is included
in subsequent quantities $N_n,V_{nz}$ and so on)

In the zero order approximation from (\ref{AB}.2) and (\ref{AB}.3)

\begin{equation}
\label{AG}
B_0=-{\beta}E_{y0}, \ \frac{1}{\gamma^2}\frac{\partial{E_{y0}}}{\partial{z}}=
\frac{\partial{E_{z0}}}{\partial{y}}
\end{equation}

From (\ref{AG}) and (\ref{AB}.4}),introducing the potential ${\varphi}_0(z,y)$
by 
\begin{equation}
\label{AH}
E_{z0}=-\frac{1}{\gamma^2}\frac{\partial{\varphi_0}}{\partial{z}}, \
E_{y0}=-\frac{\partial{\varphi_0}}{\partial{y}},
\end{equation}

it follows that

\begin{eqnarray}
\label{AI}
\frac{1}{{\gamma}^2}\frac{\partial^2{\varphi_0}}{\partial{z^2}}+
\frac{\partial^2{\varphi_0}}{\partial{y^2}}=
\left\{\begin{array}{ccc}
1, & -d \leq z \leq d, & -1 \leq y \leq b \\ 
0, & |z| >d, |y| <b
\end{array}\right.
\end{eqnarray}

In the first order approximation:
\begin{equation}
\label{AJ}
-\frac{\partial{N_1}}{\partial{z}}+\frac{\partial{N_1V_{y0}}}{\partial{y}}=0
\end{equation}

from (\ref{AJ}) and (\ref{AB}) follows:

\begin{equation}
\label{AK}
\frac{\partial^2{E_{z1}}}{\partial{z^2}}+{\gamma}^2\frac{\partial^2{E_{z1}}}
{\partial{y^2}}=\frac{\partial}{\partial{z}}\left(\frac{n_0}{n_b}-{\gamma}^2
N_1-N_1V_{z0}\right);
\end{equation}

The focusing force is

\begin{eqnarray}
\label{AL}
f_y=-e(E_y+{\beta}B)=f_{y0}+f_{y1}+..., \\ \nonumber
f_{y0}=-e(E_{y0}+{\beta}B_0)=-e{\gamma}^{-2}E_{y0}, \\ \nonumber
f_{y1}=-e(E_{y1}+{\beta}B_1) \equiv -eW_{y1}. 
\end{eqnarray}

From (\ref{AB}.3) in the first approximation

\begin{equation}
\label{AM}
\frac{\partial{W_{y1}}}{\partial{\tilde{z}}}=\frac{\partial{E_{z1}}}
{\partial{y}},
\end{equation}

which is an analog of the Panoffsky-Wenzel relation;remember that in 
(\ref{AM}) $\tilde{z}=z-v_0t$.
Differentating (\ref{AK}) over $y$ and then integrating it over $\tilde{z}$,
taking into account relation (\ref{AM}) and that \\

$\frac{\partial}{\partial{y}}\left(\frac{n_0}{n_b}-{\gamma}^2N_1-N_1V_{z0}
\right) \rightarrow 0$ when $y\rightarrow \pm \infty$,we have

\begin{equation}
\label{AN}
\frac{\partial^2{W_{y1}}}{\partial{{\bar{z}}^2}}+\frac{\partial^2{W_{y1}}}
{\partial{y^2}}=\frac{1}{\gamma^2}\frac{\partial}{\partial{y}}
\left(\frac{n_0}{n_b}-{\gamma}^2N_1-N_1V_{z0}\right)
\end{equation}

where $\bar{z} \equiv \gamma\tilde{z}$.

In what follow,we consider the ultrarelativistic bunches,when $\gamma \gg 1$;
then $f_{y0} \sim O({\gamma}^{-2})$ and in right hand side of (\ref{AN}) it is 
possible to leave only the term $-\frac{\partial{N_1}}{\partial{y}}$.
So for the sought quantity $W_{y1}$ we need $N_1$ from eq. (\ref{AJ}).Entered
in (\ref{AJ}) $V_{y0}$ must be found from eqs. (\ref{AC}),(\ref{AI}) for the 
fields in
zero approximation.It is evident,from eq. (\ref{AN}),that the role of the
noncompensated positiv ions (the term on the right hand side of eq. (\ref{AN}),
proportional to $\frac{n_0}{n_b}-N_1(1+V_{z0}))$,is negligible in 
ultrarelativistic limit ($\sim O(\gamma^{-2})$). 

\section{\small{ZERO ORDER APPROXIMATION}}

In the zero order approximation it is necessary to solve eq. (\ref{AI}) for
the flat bunch moving in vacuum ($\frac{n_0}{n_b}=0$).

The solution may be written using Green function formalism (see e.g. \cite{P})

\begin{equation}
\label{AO}
{\varphi}_0(\bar{z},y)=\frac{1}{2\pi}\int_{-\bar{d}}^{\bar{d}}d\bar{z}'
\int_{-b}^{b}dy'
\ln{\left[{(\bar{z}-\bar{z}')}^2+{(y-y')}^2\right]}^{1/2}
\end{equation}

where $\bar{d}={\gamma}d$.

Double integration in (\ref{AO}) may be performed using standart integral 
tables,but the  result of integration has a complex form,which is difficult 
to use.
It is more convinient to use (\ref{AH}) and perform an approximate integration
for $\gamma \gg 1$,taking into account,that argument of the integral has a 
complex pole,when $-d \leq z \leq d, -b \leq y \leq b$.Under these conditions
the integration over $\bar{z}$ along the real axis in complex $\bar{z}$ plane 
is equal to residue at the pole,minus integral over the semicircle with the 
radii $d$ in upper complex $\bar{z}$ half plane.The approximate results of 
integration for $\gamma \gg 1$ are the following:

\begin{eqnarray}
\label{AP}
E_{y0} \approx -y\left(1-\frac{2bd}{{\pi}{\gamma}(d^2-z^2)}\right)=
-y+O({\gamma}^{-1}) \\ \nonumber
-b \leq y \leq b,-d \leq z \leq d;
\end{eqnarray}
\begin{eqnarray}
\label{AQ}
E_{y0} \approx -b\left(1-\frac{2dy}{{\pi}{\gamma}(d^2-z^2)}\right)=
-b+O({\gamma}^{-1}) \\ \nonumber
y \geq b,-d \leq z \leq d;
\end{eqnarray}
\begin{eqnarray}
\label{AR}
E_{y0} \approx -\frac{2{\gamma}db}{{\pi}y} \\ \nonumber
y >{\gamma}d \gg b,-d \leq z \leq d;
\end{eqnarray}
\begin{eqnarray}
\label{AS}
E_{y0} \approx -\frac{2dby}{{\pi}{\gamma}z^2} \\ \nonumber
z \gg d,{\gamma}d \gg b,-b \leq y \leq b;
\end{eqnarray}
\begin{eqnarray}
\label{AT}
E_{z0}=\frac{b}{{\pi}{\gamma}}\ln\frac{(d-z)}{(d+z)},
{|y-b|}^2 \ll {\gamma}^2{|z-d|}^2 \\ \nonumber
E_{z0}=-\frac{2{\gamma}zdb}{{\pi}y^2},
{|y-b|}^2 \gg {\gamma}^2{|z-d|}^2,y>{\gamma}d \gg b
\end{eqnarray}

For comparison,remember that for the point relativistic charge

\begin{equation}
\label{AU}
E_y=E_{\bot}=\frac{e}{R}\gamma,E_z=E_{\|}=\frac{e}{{\gamma}^2R},
R^2=x^2+y^2+{(z-v_0t)}^2;
\end{equation}

i.e. the electric field for $\gamma \gg 1$ is practically transversal.In
our case of the flat bunch expression (\ref{AP}-\ref{AT}) shows that always
$|E_{z0}| \ll |E_{y0}|$,but the expressions for $E_{y0},E_{z0}$ modified,
compared to (\ref{AU}),due to the different geometry of the charge distribution.
In what follows,we will use the approximate expressions for $E_{y0}$ up to 
terms $\sim O({\gamma}^{-1})$,and put it to zero when $|z|>d$.

The next problem in zero approximation is the definition of the generalized 
velocities $V_{y0},V_{z0}$.When $\gamma \gg 1$,and bunch length is short 
enough $d' \ll 1,(d'=k_bd)$ it is possible to drop out in (\ref{AC}) the terms
proportional to
${|V_{z0}|}^2 \ll |V_{z0}|$ and to ${|V_{y0}|}^2 \ll |V_{y0}|$.
Taking into account that in the considered case $|E_{y0}| \gg |E_{z0}|$ the
equations of the motion (\ref{AC}) approximately can be rewritten in the 
following simple form

\begin{eqnarray}
\label{AV}
\frac{\partial{V_{z0}}}{\partial{z}}\approx V_{y0}E_{y0}, \\ \nonumber
\frac{\partial{V_{y0}}}{\partial{z}}\approx (1+V_{z0})E_{y0},
|y| \leq b,|z| \leq d;
\end{eqnarray}

The solutions of eqs. (\ref{AV}) for the boundary condition $V_{y0}=0,V_{z0}=0$,
when $z=d$ are:

\begin{eqnarray}
\label{AW}
V_{y0}=\sinh{(|E_{y0}|(d-z))} \simeq |E_{y0}|(d-z), \\ \nonumber
V_{z0}=\cosh{(|E_{y0}|(d-z))}-1 \simeq \frac{1}{2}{|E_{y0}|}^2{(d-z)}^2,
\end{eqnarray}

where $E_{y0}$ is given by (\ref{AP}-\ref{AR})and 
$V_{y0} \rightarrow 0,V_{z0} \rightarrow 0$ when $y \rightarrow \pm \infty$.
For the short ($d \ll 1$) bunches from (\ref{AW}) follows that $V_{z0} \ll 
V_{y0}$.

\section{\small{FIRST ORDER APPROXIMATION}}

Using eq. (\ref{AA}) in the first approximation (see (\ref{AJ}))
and eq. (\ref{AW}) for $V_{y0}$ it is possible to find $N_1(y,z)$ by the method
of characteristics.

When $|y| \leq b, E_{y0} \simeq -y+O({\gamma}^{-1})$ and characteristics are
\begin{equation}
\label{AX}
ye^{-\frac{1}{2}{(d-z)}^2}=C_1,yN_1^{(1)}=C_2.
\end{equation}
The solution of eq. (\ref{AX}), which is equal to $\frac{n_0}{n_b}$, 
when $z=d$,is

\begin{equation}
\label{AY}
N_1^{(1)}=\frac{n_0}{n_b}e^{-\frac{1}{2}{(d-z)}^2}
\end{equation}

When $|y| \geq b,E_{y0} \simeq -b+ O({\gamma}^{-1})$ and characteristic is 
$-\frac{1}{2}{(d-z)}^2 + \frac{y}{b} =C$,so $N_1^{(2)}$ is function of
$-\frac{1}{2}{(d-z)}^2+\frac{y}{b}$;
When $y=b$ it must coincides with the expression (\ref{AY}),so

\begin{equation}
\label{AZ}
N_1^{(2)}=\frac{n_0}{n_b}e^{-\frac{1}{2}{(d-z)}^2+\frac{y}{b}-1}
\end{equation}
When $y>{\gamma}d \gg b,E_{y0} \simeq -\frac{2bd\gamma}{{\pi}y}$
and characteristics are
$y^2-\frac{2bd\gamma}{\pi}{(d-z)}^2=C_1,\frac{N_1^{(3)}}{y}=C_2$

Taking into account that $N^{(3)} \rightarrow \frac{n_0}{n_b}$
when $|y| \rightarrow \infty$,we have  

\begin{equation}
\label{BA}
N_1^{(3)}=\frac{n_0}{n_b}\frac{|y|}{{\left[y^2-\frac{2bd\gamma}{\pi}
{(d-z)}^2\right]}^{1/2}}
\end{equation}

Expressions (\ref{AY}-\ref{BA})are the solutions of the eq. (\ref{AX})
for subsequent values of $V_{y0}$ (\ref{AW}).Expression (\ref{AZ}) for
$z=d$ gives
\begin{equation}
\label{BB}
N_1^{(2)}=\frac{n_0}{n_b}e^{\frac{y}{b}-1}
\end{equation}
which seems as unappropriate.Taking into account,that (\ref{AZ}) is an
approximate expression,we assume the following interpolation formula for
$N_1^{(2)}$:

\begin{equation}
\label{BC}
N_1^{(2)}=\frac{n_0}{n_b}e^{-\frac{1}{2}{(d-z)}^2+c\left(\frac{y}{b}-1\right)
{(d-z)}^2}
\end{equation}
which is $\frac{n_0}{n_b}$ when $z=d$,and coincides with the (\ref{AY}),
when $y=b$.

$N_1^{(2)}$ is exponentially rising,when $y\rightarrow+\infty$ and at the
some point $y_{cr}$ it intercept the $N_2^{(3)}$,which is also rising,when $y
\rightarrow
y_p+0 \equiv {\left(\frac{2bd\gamma}{\pi}\right)}^{1/2}(d-z)<y_{cr}$.
Hence $N_1$ has
a maximum value and the plasma electrons distribution exibits a crest at the 
point $y=y_{cr}$.

Parameters $c$ and $y_{cr}$ may be defined by the condition
\begin{equation}
\label{BD}
N_1^{(2)}(y_{cr},z)=N_1^{(3)}(y_{cr},z)
\end{equation}
The second condition may be obtained,using the following consideration:
longitudinal displacements of the plasma electrons are negligible,compared to
transverse displacements,due to condition $V_{z0} \ll V_{y0}$,so for the given
$|z| \leq d$ deficiency of the plasma electrons in the region $0<y<y_{cr}$ (and
in the symmetric region $-y_{cr}<y<0$) is equal to the surplus of the plasma 
electrons in the region $y_{cr}<y<+\infty$ (and in the region 
$-\infty<y<-y_{cr}$ subsequently):

\begin{eqnarray}
\label{BE}
\int_{0}^{b}\left(\frac{n_0}{n_b}-N_1^{(1)}\right)dy+\int_{b}^{y_c}
\left(\frac{n_0}{n_b}-N_1^{(2)}\right)dy= \\ \nonumber
=\int_{y_c}^{y_{cr}}\left(\frac{n_0}{n_b}-N_1^{(2)}\right)dy+
\int_{y_{cr}}^\infty\left(N_1^{(3)}-\frac{n_0}{n_b}\right)dy
\end{eqnarray}

In (\ref{BE}) $y_c$ is the point,where $N_{(y_c,z)}^{(2)}=\frac{n_0}{n_b}$
From (\ref{BD}) for $d \ll 1$:
\begin{equation}
\label{BF}
c=\frac{1}{2}\frac{\left(1+\frac{2bd\gamma}{{\pi}y_{cr}^2}\right)}
{\left(\frac{y_{cr}}{b}-1\right)}.
\end{equation}

From eq. (\ref{BE}) for $d\ll1$:

\begin{equation}
\label{BG}
y_{cr}=\frac{2d\gamma}{\pi} \gg b
\end{equation}

and $c$ from (\ref{BF}) approximately has a following simple form:

\begin{equation}
\label{BH}
c \approx \frac{b}{2y_{cr}}
\end{equation}

Now it is possible to turn to determination of the focusing force $W_{y1}$
from eq. (\ref{AN}). For $\gamma \gg 1$ eq. (\ref{AN}) we get an approximate 
form

\begin{equation}
\label{BI}
\frac{\partial^2{W_{y1}}}{\partial{{\bar{z}}^2}}+
\frac{\partial^2{W_{y1}}}{\partial{y^2}}\approx -\frac{\partial{N_1}}
{\partial{y}}
\end{equation}

and using the Green function formalism the solution of eq. (\ref{BI})can be 
written as
\begin{equation}
\label{BJ}
W_{y1}=-\frac{1}{2\pi}\int{d\bar{z}'}\int{dy'}\frac{\partial{N_1(y',{\bar{z}}')}}
{\partial{y'}}\ln{\left[{(y-y')}^2+{(\bar{z}-{\bar{z}}')}^2\right]}^{1/2}.
\end{equation}

The domain of the integration in (\ref{BJ}) is defined by the condition
$\frac{\partial{N_1(y,z)}}{\partial{y}} \neq 0$,i.e. it is $-\infty <y<+\infty,
-\infty<z \leq d$,because for $z>d, N_1=\frac{n_0}{n_b}$ and $\frac{\partial
{N_1}}{\partial{y}}=0$.For $-d \leq z \leq d,\frac{\partial{N_1}}{\partial{y}}$
is given by (\ref{AY}),(\ref{BA}),(\ref{BC});for $z<-d$ from the continuity
condition $N_1(z,y)=N_1(-d,y)$ up to some $-z_0<-d$ where considered steady
state regime changes to (nonlinear) wake wave (and later on the uniform 
distribution of the plasma electrons with $N_1=\frac{n_0}{n_b}$).

Due to the symmetry of the Green-function and antisymmetry of the derivative
$\frac{\partial{N_1}}{\partial{y}}$ it is possible to change the integration
over the region $-\infty <y \leq 0$ to the integration over the region
$0 \leq y \leq \infty$. Then the expression (\ref{BJ}) has the following form:

\begin{equation}
\label{BK}
W_{y1}=-\frac{1}{4\pi}\int_{-\bar{z_0}}^{\bar{d}}d{\bar{z}}'\int_b^{\infty}{dy'}
\frac{\partial{N_1(y',\bar{z}')}}
{\partial{y'}}\ln\frac{{(y-y')}^2+{(\bar{z}-\bar{z}')}^2}{{(y+y')}^2+
{(\bar{z}-\bar{z}')}^2}
\end{equation}
taking into account (\ref{AY}).The interval of the integration over $y'$ in
(\ref{BK}) must be divided in the two intervals $(b,y_{cr})$ and $(y_{cr},
\infty)$,
where eq. (\ref{BC}) and (\ref{BA}) must be used for $\frac{\partial{N_1}}
{\partial{y}}$;both expressions (\ref{BA},\ref{BC}) have a maximum at 
$y'=y_{cr}$ and it is possible approximately to take out of the integral the
slowly variing logarithmic function at the maximum point $y'=y_{cr}$.Then 
(\ref{BK}) will have the following form

\begin{equation}
\label{BL}
W_{y1} \approx -\frac{1}{4\pi}\int_{-{\bar{z_0}}}^{\bar{d}}d{\bar{z}}'\ln
\frac{{(y_{cr}-y)}^2+
{(\bar{z}-{\bar{z}}')}^2}{{(y_{cr}+y)}^2+{(\bar{z}-z')}^2}\left[\frac{n_0}{n_b}-N_1^{(2)}
(b,{\bar{z}}')\right]
\end{equation}

Notice that the approximate expression (\ref{BL}) for $W_{y1}$ does not depend
explicitely on the form of,then to some extent arbitrary,function $N_1^{(2)}$
(\ref{BC}),because $N_1^{(2)}(b,z)=N_1^{(1)}(z)$,see (\ref{BC}) and (\ref{AY}).
Only the value of $y_{cr}$ (\ref{BG}),which enters in (\ref{BL}),
explicitely depends on the form of $N_1^{(2)}$ through conditions (\ref{BD},
\ref{BE}).

Taking into account,that $|y| \leq b \ll y_{cr}$ it is possible to develop 
the logarithm function under integral in a Taylor series,leaving the first
term of the expression;then for $d \ll 1,y<b$ the expression for $W_{y1}$ 
will have the following form:

\begin{equation}
\label{BM}
W_{y1}=\frac{1}{2\pi}\frac{n_0}{n_b}y_{cr}{\gamma}y\int_{-z_0}^{d}dz'
\frac{{(d-z')}^2}{y_{cr}^2+{(z-z')}^2{\gamma}^2}
\end{equation}

Consider first the integral over $z'$ in the domain $(-d,d)$,and later on in
the domain $(-z_0,-d)$.The resulting expressions denote by $W_{y1}^{(0)}$ and
$W_{y1}^{(-)}$ subsequently;then

\begin{equation}
\label{BN}
W_{y1}^{(0)}=\frac{2\alpha(z)}{\pi}\frac{n_0}{n_b}yd^2,
\end{equation}

where $\alpha(z)$ is known function,which varies in interval $0,21 \leq 
\alpha(z) \leq 2,26$,when $-d \leq z \leq d, W_{y1}^{(-)}\approx \frac{2}{\pi}
\beta(z,z_0)\frac{n_0}{n_b}yd^2,\beta(z,z_0)$ weekly depends on the value of
the $z_0$,and it is possible to take for estimates $z_0 \rightarrow +\infty$.
Then $\beta(z)=\beta(z,\infty)$ varies in interval $0,26 \leq \beta(z)
 \leq 1,57$,when $-d \leq z \leq d$.

Hence,for estimates,it is possible to use 

\begin{equation}
\label{BO}
W_{y1}=W_{y1}^{(0)}+W_{y1}^{(-)}\approx \frac{2}{\pi}yd^2[\alpha(z)+\beta(z)]
\frac{n_0}{n_b} \approx \frac{4}{\pi}yd^2\alpha(z)\frac{n_0}{n_b}
\end{equation}

It is possible to apply the same approach to the flat,rigid,short positron 
bunch with uniform charge distribution.In this case in eqs. (\ref{AO}-\ref{AR}) 
electric field components change sign. As a consequence the $V_{0y}$
component of the generalized velocity (eq. \ref{AW}) also changes 
sign,which corresponds to attraction of the plasma electrons by positron 
bunch.Obviously enough,this leads to drastic change in plasma electrons 
redistribution as compared to electron bunch case.Instead of eq. (\ref{AY}) 
for $|y| \leq b$ it is possible to obtain

\begin{equation}
\label{A1}
N_{1p}^{(1)}=\frac{n_0}{n_b}e^{\frac{1}{2}{(d-z)}^2};
\end{equation}
instead of (\ref{AZ}) for $y \geq b$ we have
\begin{equation}
\label{A2}
N_{1p}^{(2)}=\frac{n_0}{n_b}e^{\frac{1}{2}{(d-z)}^2+y/b-1}
\end{equation}

and instead of (\ref{BA}) for $y>\gamma d \gg b$
\begin{equation}
\label{A3}
N_{1p}^{(3)}=\frac{n_0}{n_b}\frac{y}{{[y^2+\frac{2\gamma 
bd}{\pi}{(d-z)}^2]}^{1/2}};
\end{equation}

From (\ref{A1}),(\ref{A2}) it is seen that plasma electrons concentrate 
around the rear part of the positron bunch.From (\ref{A3}) it is seen that 
$N_{1p}^{(3)}$ has a complex pole and hence is slowly rising 
function,approaches it's limit equal $\frac{n_0}{n_b}$ from bellow,when 
$y \rightarrow +\infty$.This means that plasma electrons came to the 
region occupied by positron bunch, $|y| \leq b$,from the regions,where 
$y>\gamma d \gg b$.Focusing effect for the positron bunch case is caused
also by plasma electrons redistribution.But
due to different behaviour of $N_{1p}^{(2)}$ and 
$N_{1p}^{(3)}$,the interpolation between these two,which provide the main 
contribution to focusing force (\ref{AL}),(\ref{BK}),is much more 
arbitrary,than in the electron bunch case.In any choose of the 
interpolating function it must be more or less smooth,and therefore the 
estimates of the integrals like (\ref{BJ}),(\ref{BK}) will be cumbersome
enough.

These two obstacles forced to think that positron bunch case must be 
treated numerically,even in the short bunch case and in the presented 
formalism.  

\section{\small{DISCUSSION}}

The expressions for the focusing force,focusing gradient and stregth are 
subsequently

\begin{equation}
\label{BP}
f_{1y}=-eW_{1y},G=\frac{W_{1y}}{y},K=\frac{eG}{{\gamma}mc^2},
\end{equation}

$W_{1y}$ is given by (\ref{BO}),which is valid when $\gamma \gg 1$ and $d' \ll 1
(k_bd \ll 1)$.In the usual units expression (\ref{BO})has the form

\begin{equation}
\label{BQ}
W_{y1}=\frac{64\alpha(z)er_e}{\pi}(n_0n_b)(yd^2)
\end{equation}

where $r_e=e^2/mc^2$ is the electron classical radius.Subsequently,focusing
gradient and strength are:

\begin{equation}
\label{BR}
G=\frac{64\alpha(z)er_e}{\pi}(n_bn_0)d^2,K=\frac{64\alpha(z)r_e^2d^2}
{\gamma\pi}(n_bn_0)
\end{equation}
 
The expressions (\ref{BR}) differs from the previous results,which
for the flat,long enough electron bunch and underdense lens are

\begin{equation}
\label{BS}
G=4{\pi}n_0e,K=\frac{4{\pi}r_en_0}{\gamma}
\end{equation}
These are based on the understanding the focusing phenomenon as caused by the 
positive charge
of the ion column,which exists inside and behind the moving electron bunch,
because the considered bunch blows out the plasma electrons.

As it follows from the presented consideration (see,e.g.,eqs. (\ref{AN},
\ref{BI}) the net focusing effect of the ion column,caused by noncompensated
positiv charge $\frac{n_0}{n_b}-N_1(1+V_{z0})$, is proportional to 
${\gamma}^{-2}$ and is negligible in our approximation $\gamma \gg 1$,compared
to focusing effect of the plasma electrons redistribution,described by eqs.
(\ref{BI}) and (\ref{BM}).The presence of the factor $\gamma^{-2}$
in the part of the net (electric plus magnetic) focusing force,which caused by
ions,is a result of magnetic compesation of the 
subsequent focusing electric field in lab frame,where electron bunch moves
with the velocity $v_0$ and unperturbed parts of plasma are at rest.
Indeed,from  Maxwell eqs. (\ref{AB}.1),
(\ref{AB}.4) in the first approximation:

\begin{eqnarray}
\label{BT}
\frac{\partial{B_1}}{\partial{y}}={\beta}N_1V_{z0}+\beta\frac{\partial{E_{z1}}}
{\partial{z}}, \\ \nonumber
\frac{\partial{E_{z1}}}{\partial{z}}+\frac{\partial{E_{y1}}}{\partial{y}}=
\frac{n_0}{n_b}-N_1(1+V_{z0}),
\end{eqnarray}
it follows:
\begin{equation}
\label{BU}
\frac{\partial{B_1}}{\partial{y}}=-\beta\left[\frac{n_0}{n_b}-N_1(1+V_{z0})
\right]+{\beta}N_1V_{z0}-\beta\frac{\partial{E_{y1}}}{\partial{y}},
\end{equation}

The first term in right hand side of eq. (\ref{BU}),being a part of the 
displacement current entered in (\ref{BT}),represents the nonuniform 
noncompensated 
positiv ions current,which propogate with the velocity $v_0$,equal to the
velocity of the electron bunch.This term caused the compensating magnetic
field,which at $\gamma \gg 1$ cancels the focusing action of the subsequent
electric 
field.The physical reason for that - the exposed positiv
ions column effectively "moves" in plasma along with the generating electron
bunch with the velocity of it.When bunch,moving through plasma,blows out
the plasma electrons,it uncower new forward parts of the partialy 
noncompensated ion
column and this "motion" of the revealed positiv ions charge,i.e. subsequent 
current,generated compensating magnetic field.

The physical interpretation can be based also directly on the displacement
current,entered in the first eq. (\ref{BT}) through the term $\beta\frac
{\partial{E_{z1}}}{\partial{z_1}}$.Ions ahead of the bunch are completely
compensated by plasma electrons,inside the bunch ions charge is only 
partially compensated
(eq. (\ref{AY})),so exists the change of electric field component
with $\tilde{z}=z-v_0t$, which generates subsequent compensating magnetic
field.Formaly this part of displacement current coincides with the current
conditioned by "effective ion motion",as it can be seen by replacing the
electric field derivative in the first eq. (\ref{BT}) by it's value from 
Coulomb low-second eq. (\ref{BT}).It is evident that this coincidence is
based on the steady state regime,adopted at the present work.
 
In the bunch frame magnetic forces acting
on the bunch electrons are absent.Electric field of the bunch charge has no
relativistic contraction and hence the plasma electron redistribution has a 
different shape,more uniform,than in lab frame,which dimenish their focusing
effect.But the focusing effect of the plasma electrons still is proportional
to $n_b$-bunch charge density;their focusing electric field is relativistically
contracted.These factors increase the redistributed plasma electrons focusing
effect,which will be larger than focusing effect of the electric field of the 
ion column,proportional to $n_0 \ll n_b$.Lorentz transformation did not changes
the transverse component of the focusing force,so it must be the same in both
frames.Above mentioned physical arguments allow to think that it could be the
case.It is difficult to perform analytical calculations in the bunch frame due
to complexity of relativistic equation of motion of the plasma 
electrons.Recall,
that in presented calculations the plasma electrons considered as 
nonrelativistic,due to condition $(k_bd) \ll 1$.

All above mentioned general physical arguments are valid for the relativistic
bunches with arbitrary length and charge distribution.Assumptions adopted in 
the present work (short enough relativistic bunch with the uniform charge 
distribution) allows us to perform analytical calculations.

Unfortunately
the condition $k_bd \ll 1$ ($d' \ll 1$,short bunches) does not permit to use
expression (\ref{BR}) for focusing strength for the flat beams of the FFTB 
\cite{M}.In that case $n_b=7,7\cdot10^{18}cm^{-3}$ and $2,8\cdot10^{18}cm^{-3}$
and $k_b=5,2\cdot10^5cm^{-1}$ and $k_b=3,1\cdot10^{5}cm^{-1}$ consequently,
$d=2,3\cdot10^
{-2}cm$ and $k_bd=1,2\cdot10^4 \gg 1$ and $0,7\cdot10^4 \gg 1$.

Using the short 
bunches,nevertheless it is possible to obtain sufficiently strong focusing 
gradients.For example let $k_bd$ will be $k_bd=0,1$.Then $d=10^{-5}
{\left(\frac{3\cdot10^{19}}{n_b}\right)}^{1/2},G=7,5\cdot 10^{-11}n_0$ and if 
$n_0
\approx 10^{17}cm^{-3}$ as in FFTB case,$G\approx 7,5\frac{MG}{cm}$.In order 
to fullfil the condition $\frac{n_0}{n_b} \ll 1,n_b$ must be e.g.,equal 
$10^{18}cm^{-3}$,then $d=5\cdot10^{-5}cm$.May be it is possible to obtain 
such a 
short bunches experimentally,using the technique,proposed in \cite{Q},\cite{R}.

It is worthwhile to stress,that due to specific form of the redistribution of 
the plasma electrons focusing gradient $\sim e^3$ and focusing strenght 
$\sim e^4$ have a quadrupole character,due to nonuniform shape of the 
distribution of plasma electrons along the crest,which diminish,inspite of the 
factor $(n_0n_b)$,significantly the numerical values of these quantities.

This disadvantage probably can be avoided,considering the medium $k_pd 
\geq 1$,
or long $k_pd \gg 1$,bunches.The main difficulty in the analytic approach to
these cases lies in the complex form of the equations of the motion (\ref{AC}-
\ref{AE}) which are fully relativistic in these cases.Probably the medium 
and long bunch cases can be treated by computational methods only.Presented 
analytical approximate description of the short bunch case can be used then
as a physical guide and programs testing example.

It will be interesting to test experimentaly the predictions (\ref{BR}) for 
focusing gradient and strenght,particularly,the dependense of these quantities
on $n_b$ and $d$.

\section{\small{AKNOWLEDGEMENT}}

Author would like to thank A.M. Sessler,who pointed out my attention on the 
plasma lens concept,for the careful reading preliminary versions of the 
manuscript,the numerous useful suggestions and criticism.Author is indebted to
S.S. Elbakian,A.G. Khachatryan,J.S. Wurtele and S.G. Arutunian for attention,
discussion and valuable comments.

I am obliged to G. Amatuni for the significant help in preparing the manuscript
for publication.

\end{document}